\pdfoutput=1
\documentclass[twocolumn,aps,floats,amsmath]{revtex4}
\usepackage{amssymb,amsmath}
\usepackage{graphicx,float}

\begin{document}

\title{High-Temperature Dielectric Response of (1-$x$)Pb(Mg$_{1/3}$Nb$_{2/3}$)O$_3$-$x$PbTiO$_3$:\\
Does Burns Temperature Exist in Ferroelectric Relaxors?}

\author{V. Bobnar}
\affiliation{Jo\v{z}ef Stefan Institute, Jamova 39, SI-1000 Ljubljana, Slovenia}

\author{C. Filipi\v c}
\affiliation{Jo\v{z}ef Stefan Institute, Jamova 39, SI-1000 Ljubljana, Slovenia}

\author{A. Levstik}
\affiliation{Jo\v{z}ef Stefan Institute, Jamova 39, SI-1000 Ljubljana, Slovenia}

\author{Z. Kutnjak}
\affiliation{Jo\v{z}ef Stefan Institute, Jamova 39, SI-1000 Ljubljana, Slovenia}

\date{\today}

\begin{abstract}
It has been considered that polar nanoregions in relaxors form at Burns temperature $T_d\approx 600$~K. High-temperature dielectric investigations of Pb(Mg$_{1/3}$Nb$_{2/3}$)O$_3$ (PMN) and 0.7PMN-0.3PbTiO$_3$
reveal, however, that the dielectric dispersion around 600~K appears due to the surface-layer contributions. The intrinsic response, analyzed in terms of the universal scaling, imply much higher $T_d$ or formation of polar nanoregions
in a broad temperature range, while high dielectric constants manifest that polar order exists already at the highest measured temperatures of 800~K. The obtained critical exponents indicate critical behavior associated with
universality classes typically found in spin glasses.
\end{abstract}

\maketitle

A key feature of relaxors, which are usually compositionally disordered perovskites or, eventually, disordered ferroelectric polymers, is the absence of long range ferroelectric order in zero electric field at any temperature.
Relaxor state is commonly described as a network of randomly interacting polar nanoregions (PNRs), embedded in a highly polarizable medium.\cite{bs,grid,vie,dulk,bovt,vak} The formation of PNRs has been inferred
from the results of the dielectric,\cite{vie,bovt} neutron scattering,\cite{vak} and acoustic emission experiments.\cite{dulk}. It has been considered that PNRs form at a so-called Burns temperature ($T_d\approx 600$~K in
inorganic relaxors), where some physical quantities (refraction index,\cite{bs} dielectric susceptibility)\cite{grid} have been reported to deviate from the usual displacive ferroelectric behavior due to the local disorder.
However, a reinvestigation of the meaning of the Burns temperature has recently been reported \cite{tou}.

In order to investigate the dielectric behavior of relaxors around anticipated $T_d$ with high accuracy, we performed high-resolution dielectric investigations of two archetypal relaxor systems, Pb(Mg$_{1/3}$Nb$_{2/3}$)O$_3$ (PMN)
single crystal and (1-$x$)PMN-$x$PbTiO$_3$ ($x=0.3$, denoted as PMN-30PT) ceramics, in the temperature range from room temperature to 800~K. Contrary to other experimental methods, which detect the polarization indirectly
and are thus more vulnerable to artefacts, for example, coupling to the elastic stress fields, dielectric spectroscopy probes directly the fluctuation of the polar order and is thus the main method which should be able to provide information
on the PNRs formation. Opposite of expectations, no anomaly which could be attributed to the formation of PNRs has been detected. We will show that dielectric dispersion at lower frequencies is due to the surface-layer contributions,
while the intrinsic data can be fitted to the universal scaling ansatz in a broad temperature range, and that such an analysis is much more reliable than using a traditional mean-field approach.

For dielectric measurements, surfaces of the [001]-oriented PMN crystal and PMN-30PT ceramic sample were covered by sputtered platinum electrodes. The complex dielectric constant
$\varepsilon^{*}=\varepsilon'-i\varepsilon''$ was measured by Novocontrol Alpha High Resolution Dielectric Analyzer in a frequency range of 0.01~Hz--10~MHz. The amplitude of the probing ac electric signal, applied to
samples with thicknesses of 0.5--1~mm, was 1~V. The temperature was stabilized by Novotherm-HT 1400 high-temperature control system. The data were obtained either
during zero-field-cooling (ZFC) or zero-field-heating (ZFH) runs, or during zero-field-heating run after the sample had been cooled in a dc bias electric field (ZFH/FC), with the cooling/heating rates between 0.15-0.50~K/min.

\begin{figure}
\centering
\includegraphics[clip,angle=0,width=7.5cm]{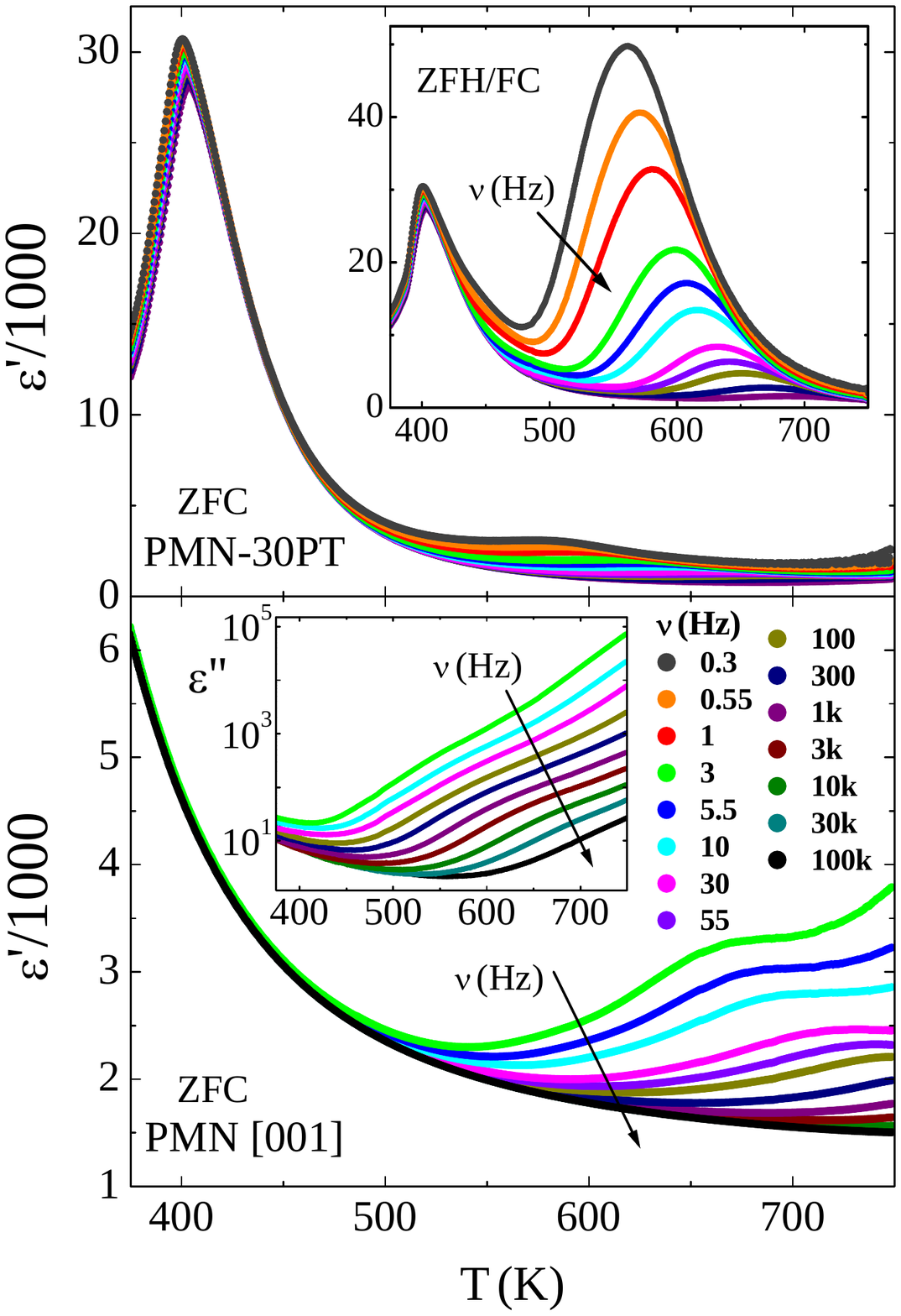}
\caption{Dielectric response of the PMN-30PT ceramics and PMN single crystal, obtained during ZFC runs with the rate of 0.35~K/min. The upper inset shows the dispersion around 600~K after the sample had been cooled in a dc bias field.}
\label{extrinsic}
\end{figure}

Figure\ \ref{extrinsic} shows the dielectric constant $\varepsilon'$ in the PMN-30PT ceramics and PMN single crystal, measured at various frequencies during ZFC runs. A typical relaxor maximum appears
in the PMN at temperatures below 300~K,\cite{lev} while it can clearly be seen in the PMN-30PT around 400~K. It should be stressed that, contrary to the PMN which undergoes the transition into a long-range ferroelectric
phase only in a dc bias field which is higher than some critical field, the PMN-30PT system undergoes a spontaneous relaxor-to-ferroelectric phase transition at a temperature slightly below the dynamic relaxor maximum.\cite{xu} In
addition, at higher temperatures, around 600~K, a significant dielectric anomaly was detected. This dispersion appears due to the Maxwell-Wagner-type contributions of interface layers between sample and contacts. This effect
has already been detected in various perovskite systems\cite{stu,bid} and is especially well known in semiconductors, where depletion layers are formed in the region close to the metallic electrodes. Then,
below a characteristic frequency of $\approx 1/RC_i$ ($R$ is the intrinsic sample resistance and $C_i$ the capacitance of interface layers) the measured capacitance is dominated by large $C_i$.\cite{lunk} While in semiconducting samples
this increase usually appears within the measured frequency window, in ferroelectrics this deviation would occur in $\mu$Hz region, as their resistance is typically much higher. However, at higher temperatures $R$ decreases
enough ($R_{dc}\approx 250$~k$\Omega$ in the PMN-30PT sample at 770~K) that already data in our frequency window are affected. This effect alters, however, only low-frequency data. In the PMN there is no anomaly in the data
measured at 100~kHz, while in the PMN-30PT no deviations were detected above 1~kHz. Additional experiments confirmed the extrinsic origin of this dispersion: (i) it strongly increases after the sample had been cooled in a dc
bias field (the upper inset to Fig.\ \ref{extrinsic}, $E_{dc}=3$~kV/cm), (ii) the sample geometry affects the dispersion (the absolute value of $R$ determines the characteristic frequency), (iii) the dispersion was strongly influenced by
using different (InGa) electrodes.

\begin{figure}
\centering
\includegraphics[clip,angle=0,width=7.5cm]{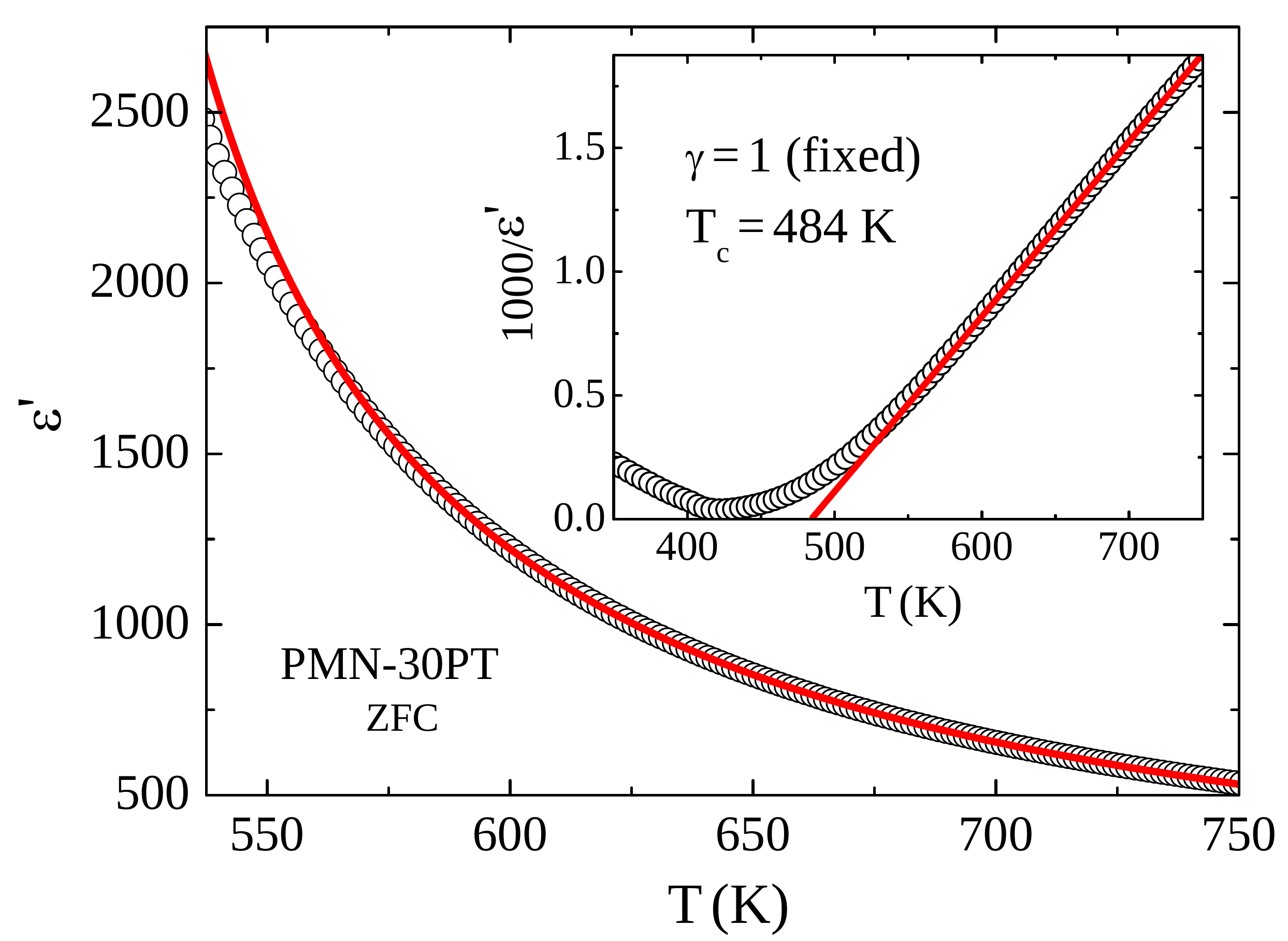}
\caption{$\varepsilon'$ in the PMN-30PT ceramics, detected during ZFC run at 100 kHz. The solid line is the fit to the mean-field behavior. The same results are shown in the inset in the form which is commonly used to linearize the data.}
\label{mf}
\end{figure}
\begin{figure}
\centering
\includegraphics[clip,angle=0,width=7.5cm]{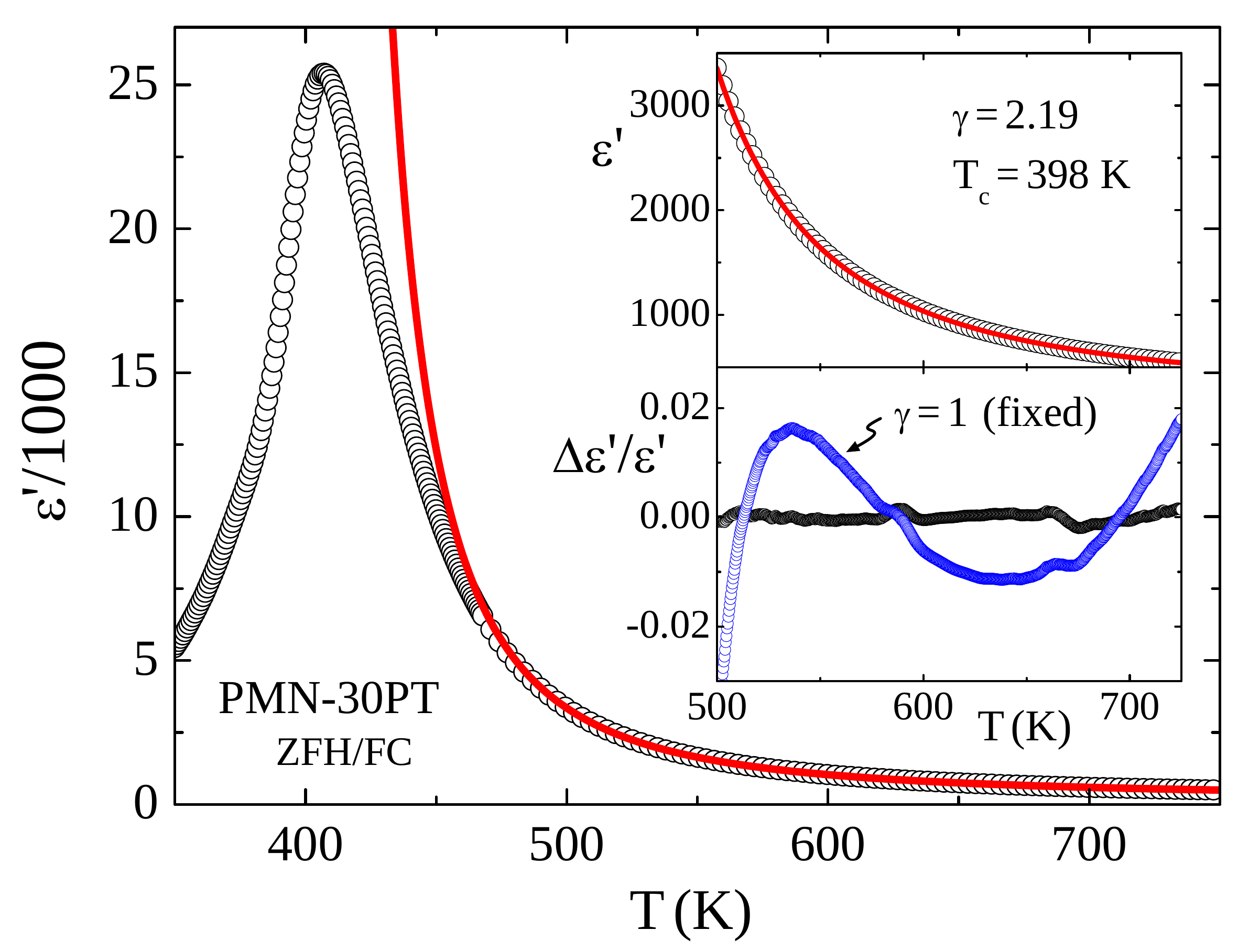}
\caption{$\varepsilon'$ vs. temperature in the PMN-30PT. The solid line is the fit to Eq. (1). The upper inset shows the data in the temperature range, where experimental data have actually been fitted. The lower inset reveals that there
is no systematic deviation of the fit from the experimental data.}
\label{cb}
\end{figure}
\begin{figure}
\centering
\includegraphics[clip,angle=0,width=7.5cm]{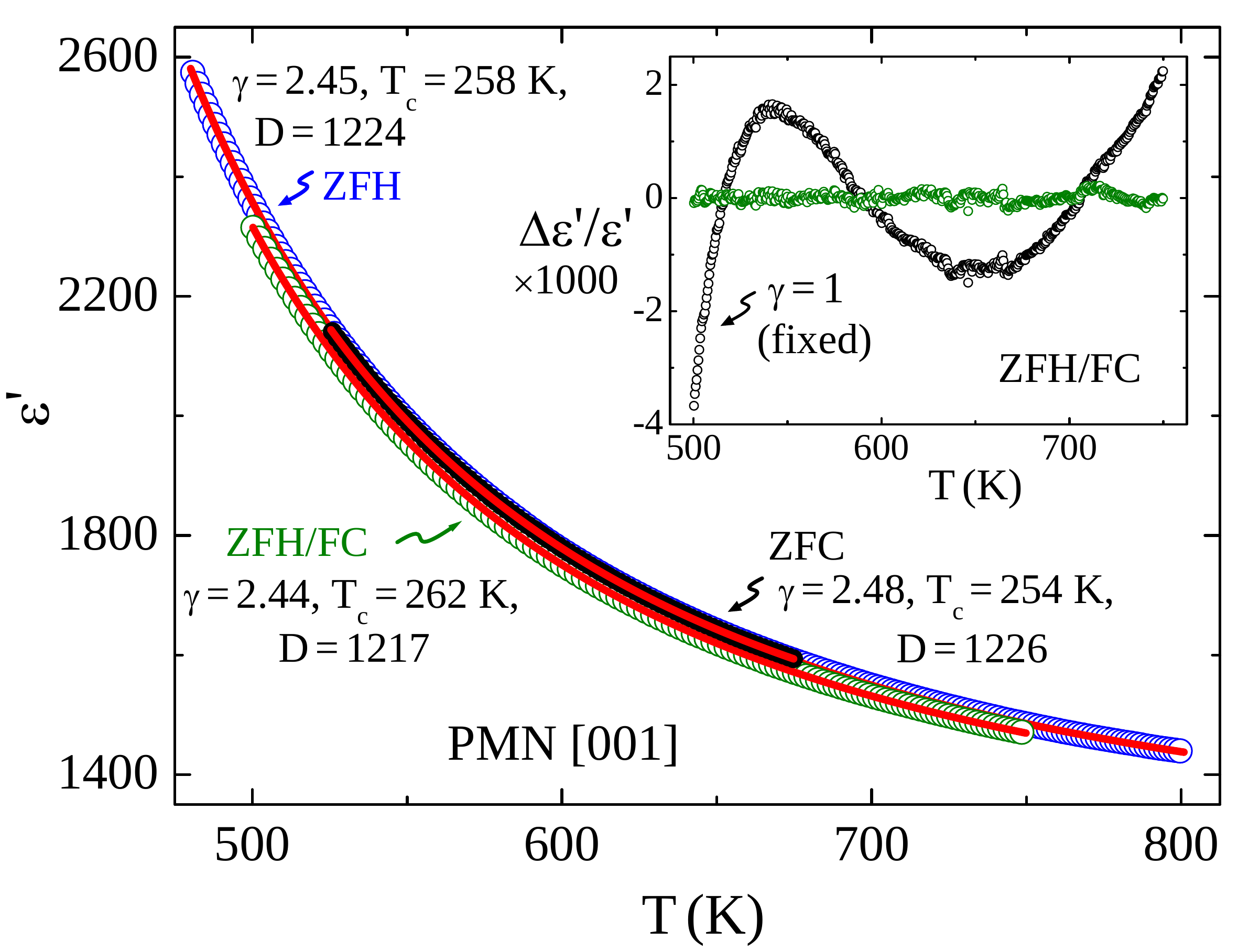}
\caption{$\varepsilon'$ vs. temperature in the PMN, detected during ZFC, ZFH, and ZFH/FC ($E_{dc}=6.5$~kV/cm) runs. Solid lines are fits to the universal scaling ansatz. The inset shows that, contrary to the mean-field fit,
there is no systematic deviation of the fit from the experimental data.}
\label{fits}
\end{figure}

The intrinsic dielectric data in the PMN-30PT were fitted to the mean-field and universal scaling ansatzes. The solid line in Fig.\ \ref{mf} is the fit to the mean-field behavior 
$\varepsilon'=C/\tau +D$ ($\tau=(T-T_c)/T_c$ and D is background), performed in the range of 625--750~K and plotted also at lower temperatures. Note that due to clarity not all experimental points are shown in
Figs.\ \ref{mf}--\ref{fits} -- the actual number of data points fitted is depicted in Table I. The same data are also shown in the inset, in the form which is commonly used to linearize the data at higher temperatures. Namely, one way
to determine $T_d$, found in literature, is to observe the temperature at which the mean-field fit starts to deviate from experimental points.  Such approach
gives  $T_d\approx 600$~K in our case. Although the fit appears to be credible, there is a systematic deviation from the experimental data, as is demonstrated in Fig.\ \ref{cb}. The fitting parameters also strongly depend on
fitting range and the fit yields an unreasonable $T_c=484$~K.

\begin{table*}
\caption{\label{tab:table1}Least-square values of the fitting parameters appearing in Eq. 1. Quantities enclosed in square brackets were held fixed during a fit: $\gamma=1$ and $a_{\varepsilon}=0$ thus denote mean-field fit.
For all fits the correction-to-scaling exponent was $\Delta=0.5$.}
\begin{ruledtabular}
\begin{tabular}{cccccccccc}
sample & run & range (K) & N$^{\circ}$ of points & $\gamma$ & $T_c$ (K) & $C$ & $a_{\varepsilon}$ & $D$ & ${\chi}_{\nu}^2$\\
\hline
PMN & ZFH & 480-800 & 778 & 2.45 & 258 & 241 & 3.07 & 1224 & 1.03 \\
PMN & ZFC & 525-675 & 272 & 2.48 & 254 & 235 & 3.44 & 1226 & 1.02 \\
PMN & ZFH/FC & 500-750 & 496 &  2.44 & 262 & 214 & 3.20 & 1217 & 1.01 \\
PMN & ZFH/FC & 500-750 & 496 & [1] & 376 & 426 & [0] & 1038 & 231 \\
\hline
PMN-30PT & ZFH/FC & 500-725 & 1302 & 2.19 & 398 & 94 & 1.35 & 221 & 1.02 \\
PMN-30PT & ZFH/FC & 500-725 & 1302 & [1] & 455 & 349 & [0] & -57 & 283 \\
\end{tabular}
\end{ruledtabular}
\end{table*}

Figure\ \ref{cb} presents the analysis of the dielectric data in terms of the universal scaling. It shows $\varepsilon'$ as a function of the temperature in the PMN-30PT ceramics, detected during ZFH/FC run (applied
dc field during cooling was 10 kV/cm), and the fit to the universal scaling ansatz
\begin{equation}
\varepsilon'=C{\tau}^{-\gamma}(1+a_{\varepsilon}{\tau}^{\Delta})+D.
\label{scal}
\end{equation}
This expression, besides the asymptotic behavior takes into account also correction to scaling due to the fluctuations of the order-parameter.\cite{see} The expansion in the brackets, commonly known as the Wegner expansion,\cite{weg}
has been derived within renormalization-group theories as for magnetic\cite{kaul} as well as for the dielectric susceptibility,\cite{bely} and contains the leading system-independent amplitude $a_{\varepsilon}$ and an additional
universal exponent $\Delta\cong0.5$. The upper inset to Fig.\ \ref{cb} shows the data in the temperature range, where experimental data have actually been fitted. Below 500 K, due to smearing effects, being common in disordered
systems, the experimental data deviate from the critical behavior, as can be seen in the main frame. A residuum plot (lower inset, $\Delta\varepsilon'=\varepsilon'_{\rm{exp}}-\varepsilon'_{\rm{fit}}$) reveals that, contrary to the
mean-field fit, there is no systematic deviation of the fit from the experimental data. This fit yields the actual $T_c$. It is noteworthy to point out that the obtained value of  $\gamma$ is very close to values predicted for 3D Ising
spin glass with binomial near-neighbor interactions,\cite{mar} and has already experimentally been obtained in an insulating spin glass.\cite{vin}

Figure\ \ref{fits} shows the intrinsic data obtained during ZFC, ZFH, and ZFH/FC ($E_{dc}=6.5$~kV/cm) runs  in the PMN single crystal. Solid lines through all three data sets are fits to Eq. 1. Similar to the PMN-30PT,
there is no systematic deviation of the fit from the experimental data, contrary to the best mean-field fit. Reliability of fits is demonstrated with their high precision (see residuums in the inset) and with the fact that  values of the fitting
parameters remain unchanged in different measurement runs and are insensible to the fitting range. This holds true as for $\gamma$ as for the values of the background and attempted $T_c$ (which is close to the critical temperature
of the induced ferroelectric transition),\cite{kut} and is clearly depicted in Table I, where values of all fitting parameters for both systems are summarized. Based on the normalized chi-square values ${\chi}_{\nu}^2={\chi}^2/{\sigma}^2$
the mean-field results can be rejected on the confidence level better than 99 \% according to the F-test (it is interesting to note that strong deviation from the mean-field behavior has just recently been found also in a ferroelectric
PbZr$_{0.5}$Ti$_{0.5}$O$_3$).\cite{alm} While in the polycrystalline PMN-30PT sample the obtained $\gamma$ is only an effective critical coefficient, here, in the single crystal, stable $\gamma$ values in fact demonstrate that
relaxors adopt critical behavior similar to that observed in 3D spin glass universality classes.\cite{mar} While dielectric investigations performed at lower temperatures revealed that relaxors in low-external electric field can be at least
empirically described as the dipolar glass state,\cite{lev,nl} high-temperature investigations manifest that the glasslike fingerprint behavior can be observed well above the dispersive dielectric maximum.

In conclusion, we show that no formation of PNRs takes place in the PMN and PMN-30PT systems below 800 K. Any sharp or smeared anomaly related to the formation of PNRs would namely be detected as a systematic deviation at
$T_d$ in our residuum plots. Dielectric investigations clearly reveal that dielectric dispersion, detected in both systems around 600~K, is due to the Maxwell-Wagner-type surface-layer contributions. The intrinsic ZFC, ZFH, and ZFH/FC
dielectric data can in both systems be reliably fitted to the universal scaling ansatz in a broad temperature range of 500--800~K. The obtained values of critical exponents indicate that relaxors adopt critical behavior associated with
universality classes typically found in spin glasses. Our results imply that either $T_d$ is above 800~K or that PNRs are continuously formed in  a broad temperature range. High values of the dielectric constant (over 1000 in the PMN)
indicate, however, that the start of PNRs' formation must take place at temperatures well above 800 K.

This research was supported by the Slovenian Research Agency under program P1-0125.


\begin{thebibliography}{99}

\bibitem{bs} G. Burns and B. A. Scott, Solid State Commun. {\bf 13}, 423 (1973).

\bibitem{vie} D. Viehland, S. J. Jang, L. E. Cross, and M. Wuttig, Phys. Rev. B {\bf 46}, 8003 (1992).

\bibitem{bovt} V. Bovtun, J. Petzelt, V. Porokhonskyy, S. Kamba, and Y. Yakimenko, J. Eur. Ceram. Soc. {\bf 21}, 1307 (2001).

\bibitem{vak} S. B. Vakhrushev and S. M. Shapiro, Phys. Rev. B {\bf 66}, 214101 (2002). 

\bibitem{dulk} E. Dulkin, I. P. Raevskii, and S. M. Emelyanov, Phys. Solid State {\bf 45}, 158 (2003).

\bibitem{grid} S. A. Gridnev, A. A. Glazunov, and A. N. Tsotsorin, phys. stat. sol. (a) {\bf 202}, R122 (2005).

\bibitem{tou} J. Toulouse, Ferroelectrics  {\bf 369}, 203 (2008).

\bibitem{lev} A. Levstik, Z. Kutnjak, C. Filipi\v c, and R. Pirc, Phys. Rev. B {\bf 57}, 11204 (1998).

\bibitem{xu} G. Xu, D. Viehland, J. F. Li, P. M. Gehring, and G. Shirane, Phys. Rev. B {\bf 68}, 212410 (2003).

\bibitem{stu} R. Stumpe, D. Wagner, and D. B\"{a}uerle, phys. stat. sol. (a) {\bf 75}, 143 (1983).

\bibitem{bid} O. Bidault, P. Goux, M. Kchikech, M. Belkaoumi, and M. Maglione, Phys. Rev. B {\bf 49}, 7868 (1994).

\bibitem{lunk} P. Lunkenheimer {\it et al.}, Phys. Rev. B {\bf 66}, 052105 (2002).

\bibitem{see} M. Seeger, S. N. Kaul, H. Kronm\"{u}ller, and R. Reisser, Phys. Rev. B {\bf 51}, 12585 (1995).

\bibitem{weg} F. J. Wegner, Phys. Rev. B {\bf 5}, 4529 (1972).

\bibitem{kaul} S. N. Kaul, Phys. Rev. B {\bf 38}, 9178 (1988). 

\bibitem{bely} M. Y. Belyakov and S. B. Kiselev, Physica A {\bf 190}, 75 (1992).

\bibitem{mar} P. O. Mari and I. A. Campbell, Phys. Rev. B {\bf 65}, 184409 (2002).

\bibitem{vin} E. Vincent and J. Hammann, J. Phys. C: Solid State Phys. {\bf 20}, 2659 (1987).

\bibitem{kut} Z. Kutnjak, B. Vodopivec, and R. Blinc, Phys. Rev. B {\bf 77}, 054102 (2008).

\bibitem{alm} E. Almahmoud, I. Kornev, and L. Bellaiche, Phys. Rev. Lett. {\bf 102}, 105701 (2009).

\bibitem{nl} V. Bobnar, Z. Kutnjak, R. Pirc, R. Blinc, and A. Levstik, Phys. Rev. Lett. {\bf 84}, 5892 (2000).

\end{thebibliography}
\end{document}